\documentclass[pra,aps,superscriptaddress,amsmath,amssymb,twocolumn,reprint]{revtex4-2}
\usepackage{amsmath,amssymb,amsfonts,braket,graphicx,float,subfigure}
\usepackage{CJKutf8}
\usepackage{bm} 
\usepackage{calc}
\usepackage{units} 
\usepackage{lipsum}
\usepackage{tikz}
\usepackage{caption}
\usepackage{ulem}
\usepackage{float}
\def\be{\begin{equation}}
\def\ee{\end{equation}}
\def\ba{\begin{eqnarray}}
\def\ea{\end{eqnarray}}
\def\sdg{Schr\"odinger~}
\def\md{\mathrm{d}}

\usepackage{dcolumn}
\usepackage{mathptmx} 
\usepackage{mathrsfs}
\usepackage{verbatim}
\usepackage{xcolor}
\definecolor{myblue}{RGB}{46,48,146}
\usepackage[colorlinks=true,linkcolor=myblue,anchorcolor=red,citecolor=myblue, urlcolor=myblue]{hyperref}
\usepackage{pifont}

\makeatletter
\renewcommand*\env@matrix[1][\arraystretch]{%
	\edef\arraystretch{#1}%
	\hskip -\arraycolsep
	\let\@ifnextchar\new@ifnextchar
	\array{*\c@MaxMatrixCols c}}
\makeatother

\newlength{\seplinewidth}
\newlength{\seplinesep}
\colorlet{sepline}{orange}

\begin{document}
\begin{CJK*}{UTF8}{gbsn}
\title{Quantum Hamiltonian Algorithms for Maximum Independent Sets}

\author{Xianjue Zhao(赵贤觉)}
\affiliation{International Center for Quantum Materials, School of Physics, Peking University, Beijing 100871, China}
\author{Peiyun Ge(葛培云)}
\affiliation{State Key Laboratory of Low Dimensional Quantum Physics, Department of Physics, Tsinghua University, Beijing 100084, China}
\author{Hongye Yu(余泓烨)}
\affiliation{Department of Physics and Astronomy, Stony Brook University, Stony Brook, NY 11794, USA}
\author{Li You(尤力)}
\affiliation{State Key Laboratory of Low Dimensional Quantum Physics, Department of Physics, Tsinghua University, Beijing 100084, China}
\affiliation{Frontier Science Center for Quantum Information, Beijing, China}
\affiliation{Beijing Academy of Quantum Information Sciences, Beijing 100193, China}
\author{Frank Wilczek}
\affiliation{Center for Theoretical Physics, MIT, Cambridge, Massachusetts 02139, USA}
\affiliation{T. D. Lee Institute and Wilczek Quantum Center, Shanghai Jiao Tong University, Shanghai 200240, China}
\affiliation{Department of Physics, Stockholm University, Stockholm SE-106 91, Sweden}
\affiliation{Department of Physics, Arizona State University, Tempe, Arizona 25287, USA}

\author{Biao Wu(吴飙)}
\email{wubiao@pku.edu.cn}
\affiliation{International Center for Quantum Materials, School of Physics, Peking University, Beijing 100871, China}
\affiliation{Wilczek Quantum Center, University of Science and Technology of China,
Shanghai 201315, China}
\affiliation{Hefei National Laboratory, Hefei 230088, China}

\date{\today}

\begin{abstract}
With qubits encoded into atomic ground and Rydberg states and situated on the vertexes of a graph, the conditional quantum dynamics of Rydberg blockade, which inhibits simultaneous excitation of nearby atoms, has been employed recently to find maximum independent sets following an adiabatic evolution algorithm hereafter denoted by HV [Science \textbf{376},  1209 (2022)]. An alternative algorithm, short named the PK algorithm, reveals that the independent sets diffuse over a media graph governed by a non-abelian gauge matrix of an emergent PXP model. This work shows the above two algorithms are {\it mathematically\/} equivalent, despite of their seemingly different physical implementations. More importantly, we demonstrated that although the two are mathematically equivalent, the PK algorithm exhibits more efficient and resource-saving performance. Within the same range of experimental parameters, our numerical studies suggest that the PK algorithm performs at least 25\% better on average and saves at least $6\times10^6$ measurements ($\sim 900$ hours of continuous operation) for each graph when compared to the HV algorithm. We further consider the measurement error and point out that this may cause the oscillations in the performance of the HV's optimization process.
\end{abstract}

\maketitle
\end{CJK*}

\begin{figure}[htp]
  \centering
  \begin{tikzpicture}
  \coordinate (1) at (0,0);
  \coordinate (2) at (1.5,0);
  \coordinate (3) at (3,0);
  \coordinate (4) at (0,-1);
  \coordinate (5) at (1.5,-1);
  \coordinate (6) at (3,-1);
  \coordinate (7) at (0.75,-2);
  \coordinate (8) at (2.25,-2);
  \draw (1)--(2);
  \draw (2)--(3);
  \draw (1)--(4);
  \draw (2)--(5);
  \draw (3)--(6);
  \draw (4)--(5);
  \draw (6)--(5);
  \draw (4)--(7);
  \draw (5)--(7);
  \draw (8)--(5);
  \draw (8)--(6);
  \draw (7)--(8);
  \fill[red] (2) circle (2pt);
  \fill[red] (4) circle (2pt);
  \fill[red] (6) circle (2pt);
  \fill (1) circle (2pt);
  \fill (3) circle (2pt);
  \fill (5) circle (2pt);
  \fill (7) circle (2pt);
  \fill (8) circle (2pt);
  \end{tikzpicture}
  \caption{{A graph with 8 vertices and 12 edges.} The circles stands for vertices and the lines stands for edges. The red circles form one of its maximum independent sets.}
  \label{fig:is}
\end{figure}
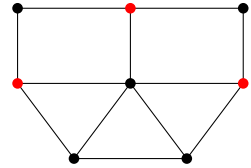

{\it Introduction.}---An independent set (IS) of a graph is a collection of vertices, none of which are directly connected by edges.  Among all the independent sets, the one with the largest number of vertices is called the maximum independent set (MIS). The red colored circles in Fig. \ref{fig:is} illustrates a MIS for a graph with 8 vertices and 12 edges. Finding MIS is a NP-hard problem on classical computer~\cite{strong}. Because of the broad prospective applications enabled by MIS, from logistics and supply chain optimization \cite{Hayes,Bodino,Employeescheduling}, to possible mapping into other NP-hard problems \cite{Garey, Bondy}, interests in efficient and effective MIS solutions are high, especially since the first reported experimental MIS solution in an Rydberg atom quantum simulator \cite{Lukin2022}. The best classical algorithm, discovered by Mingyu Xiao \cite{Xiao}, reaches a scaling of $1.1996^{n}n^{O(1)}$ with $n$ being the number of vertices characterizing the size of a graph. 

Recently, two apparently different quantum algorithms have been proposed for MIS \cite{Lukin2022,wupra}. While it remains unclear whether they 
might offer any quantum advantage over classical algorithms, the promising scalability of the experimental platform employed in \cite{Lukin2022,Lukin2023,Lukin2024} raises significant hope that it may provide a viable experimental approach for finding MIS of $n>2000$. 

Both algorithms are referenced to the transverse-Ising type Hamiltonian with nearly the same forms for atom-atom interactions. As we shall describe below, the blockade mechanism between nearby atoms can be tuned to inhibit Rydberg atoms from appearing in connected vertices. The HV algorithm carefully optimizes one-body operators as detailed in \cite{Lukin2022}. The PK algorithm was introduced and refined in \cite{wupra,yucpl}, and its evolution can be understood in terms of non-abelian gauge potentials during adiabatic state evolution.

This Letter reports enhanced understanding gained for MIS in a graph of Rydberg atoms. First, the physical implementations behind the above two referenced algorithms are found mathematically equivalent: the PK algorithm is the HV algorithm transformed into the interaction picture. This equivalence reveals a profound connection to many-body quantum dynamics. The PXP model~\cite{Turner,Turner2,Turner3,Serbyn,Lei}, known for its quantum many-body scar phenomenon and experimentally realized with Rydberg atom arrays \cite{Lukin2017}, is essentially the non-abelian gauge matrix in the PK algorithm. 
Realization of this relationship gives us new perspectives. The PXP model, traditionally viewed as a many-body problem, can now be seen as single-particle quantum diffusion over a median graph~\cite{Schaefer,Chepoi,wiki2sat}, whose vertices represent the independent sets of the original graph. Moreover, the PXP model, originally defined on a special 
one-dimensional graph, can be extended to other graphs. 

Second, the approach based on the PK algorithm is more effective and requires fewer measurements in experiment. In numerical simulations, we find the analytic variational ansatz of the PK algorithm to the adiabatic path outperforms the brute-force segment-wise numerically optimized path in HV algorithm by at least 25\% and saves at least $6\times10^6$ measurements ($\sim 900$ hours of continuous operation) for each graph on average. Such advantage in efficiency is found to persist for various graph types and for different sizes, raising the heightened desire for future experiments with larger sized tweezer arrays~\cite{Endres, Sachdev, SaffmanRMP, Theis,Zeng,Levine, Cohen, You, Youprr, Cong}. 

{\it Quantum algorithms.}---The HV algorithm \cite{Lukin2022} is referenced to the following Hamiltonian
\begin{eqnarray}
\hat{H}_{\text{HV}}&=&\frac{\hbar}{2}\sum_{j=1}^{n}[\Omega(t)e^{i\varphi(t)}|0\rangle_j\langle 1|_{j} + {\rm h.c.} -2\Delta(t)\hat{n}_j ]  \nonumber \\
\ \ \ \ \ &{}& +\sum_{i<j}V_{ij}\hat{n}_i\hat{n}_j\,,
\label{1}
\end{eqnarray}
where $|0\rangle_j$ represents that the atom at site $j$ not in the excited Rydberg state, while $|1\rangle_j$ represents that it is excited. For a graph with $n$ vertices of single atoms, the operator $\hat{n}_j = |1\rangle_j \langle 1|_j$ denotes the Rydberg state fraction operator for site $j$. The term $V_{ij}$ describes the interaction strength between two excited Rydberg atoms situated at sites $i$ and $j$.  A repulsive interaction $V_{ij}$ imposes an energy penalty on multi-atom configurations in which both $i$ and $j\not = i$ are excited.  Such interactions correspond, in the graph problem, to the existence of a line connecting the sites.   $\Omega(t)$, $\varphi(t)$, and $\Delta (t)$ are control functions whose implementations define the algorithm.  In general, the goal of MIS algorithms of the type discussed here is to evolve into configuration with many disconnected excited states.  These correspond to low-energy states at late times, when $\Omega \rightarrow 0$ and $\Delta$ approaches a positive constant.

With the pseudo-spin operators $\sigma_j^z=2\hat{n}_j-1$, 
$\sigma_j^+=|1\rangle_j\langle 0|_j$, and $\sigma_j^-=|0\rangle_j\langle 1|_j$ we can rewrite the Hamiltonian $\hat{H}_{\text{HV}}$ (up to an irrelevant time-dependent c-number) as $
\hat{H}_{\text{HV}}=\hat{H}_1(t)+\hat{H}_2$, where 
\begin{eqnarray}
  \label{h1}
\hat{H}_1(t)&=&\frac{\hbar}{2}\Omega(t)\cos{\varphi(t)}\sum_j \sigma_j^x+\frac{\hbar}{2}\Omega(t)\sin{\varphi(t)}\sum_j \sigma_j^y \nonumber\\
\ \ \ \ \ &{}&
-\frac{\hbar}{2}\Delta(t)\sum_j \sigma_j^z\,,
\end{eqnarray}
and $\hat{H}_2=\sum_{\langle i,j\rangle}V_{ij}\hat{n}_i\hat{n}_j$. Here $\hat{H}_{\text{HV}}$ is partitioned into two parts: $\hat{H}_1(t)$, a single-spin Hamiltonian which depends on time, and $\hat{H}_2$,  
the interactions between spins which is time-independent.

The PK algorithm in \cite{wupra,yucpl} proposed  theoretically a seemingly different Hamiltonian
\be
\label{ha}
\hat{H}_{\text{PK}}(t) = U(t)\hat{H}_2'U^\dagger(t)\,,
\ee
where $\hat{H}_2'= \sum_{\langle i,j\rangle}V_0\hat{n}_i\hat{n}_j$, and $U(t)=V(t)^{\otimes n}$ with $V(t)$ being a unitary matrix for spin-1/2. When $V(t)$ changes adiabatically as specified \cite{wupra,yucpl}, one finds the MISs as the ground states  according to the PK algorithm. 

{\it Equivalence.}---We now demonstrate the Hamiltonians $\hat{H}_{\text{HV}}$ and $\hat{H}_{\text{PK}}$ are theoretically equivalent. The starting point is to note that $\hat{H}_2'$ and $\hat{H}_2$ are essentially the same for $V_{ij} > 0$, or repulsive interaction between Rydberg atoms. This is because one can always choose a $V_0 > 0$ such that all $V_{ij} > V_0$. In this case, $\hat{H}_2'$ and $\hat{H}_2$ have the same set of ground states, which correspond to all independent sets of a given graph. Therefore, only $\hat{H}_2$ will be referenced to in the following discussion.

Consider $\hat{H}_{\text{HV}} = \hat{H}_1(t) + \hat{H}_2$, the evolution of its wave function $\ket{\Phi_S(t)}$ is described by the \sdg equation $
i\md \ket{\Phi_S(t)}/\md t=
[\hat{H}_1(t)+ \hat{H}_2]\ket{\Phi_S(t)}
$. We can go to the interaction picture with the following unitary evolution operator $
U_I(t) = \mathcal{T}e^{-i\int_0^t\hat{{H}}_1(t')\mathrm{d}t'}$, and the quantum state $\ket{\Phi_I(t)}$ in the interaction picture is related to the state in the \sdg picture as follows
$\ket{\Phi_I(t)}=U_I^\dagger(t)\ket{\Phi_S(t)}$, which satisfies the following equation $
i\md \ket{\Phi_I(t)}/\md t=
U_I^\dagger(t)\hat{H}_2 U_I(t)\ket{\Phi_I(t)}$.

When $U_I(t)=U^\dagger(t)$, it is clear that the state $\ket{\Phi_I(t)}$ will follow the evolution governed by the Hamiltonian $\hat{H}_{\text{PK}}(t)$ as specified 
in Eq. (\ref{ha}).  With the form of $U(t)$ given in \cite{wupra,yucpl}, the condition 
$U_I(t)=U^\dagger(t)$ allows us to deduce  $\hat{H}_{1}(t)$.
If $\hat{H}_{1}(t)$ takes the same form of  $\hat{H}_{1}(t)$ in Eq. (\ref{h1}), 
the Hamiltonian $\hat{H}_{\text{PK}}(t)$ is equivalent to $\hat{H}_{\text{HV}}(t)$.

After some calculations, we obtain
\begin{equation}
  \hat{H}_1 =-i U^\dagger(t)\partial_t U(t)=(\vec{\mu}\times\vec{\mu}')\cdot\sum_j\vec{\sigma}_j\,.
  \label{h1new}
\end{equation}
where $\vec{\mu}(t)=\left(\sin (\theta/2) \cos\phi, \sin (\theta/2) \sin\phi, \cos (\theta/2)\right)$ and $\vec{\mu}'(t)=\md\vec{\mu}(t)/\md t$, with $\theta$ and $\phi$ changing with time according to $\theta = \omega_\theta t$ and $\phi = \omega_\phi t\,$. The physical meaning of $\theta$ and $\phi$ can be found in \cite{wupra,yucpl}.

The $\hat{H}_{1}(t)$ in Eq. (\ref{h1new}) is of the same form of  $\hat{H}_{1}(t)$ in Eq. (\ref{h1}). We thus have 
shown that the Hamiltonian $\hat{H}_{\text{PK}}(t)$ is equivalent to $\hat{H}_{\text{HV}}(t)$.

In the HV experiment \cite{Lukin2022}, the quantum state is in the \sdg picture, which as we show above is related to the state in the interaction picture according to $\ket{\Phi_I(t)}=U_I^\dagger(t)\ket{\Phi_S(t)}$. At the end $t=T$ of the adiabatic evolution  specified in Ref. \cite{yucpl}, we have $\theta=\pi$ and $\phi=0$. 
This suggests that $U_I^\dagger(T)= U(T) = [\sigma_x ]^{\otimes n}$. Its action is to flip every qubit. For the PK algorithm as discussed in Ref. \cite{yucpl}, one needs to flip all the spins to get the right answer for the MIS. So the two flips cancel out and we can determine MIS from Rydberg atom distribution associated with $\ket{\Phi_S(t)}$ finally obtained in the experiment.

{\it Non-abelian gauge potential and the PXP model.}---We note that $-\hat{H}_1(t)$ is actually a non-abelian gauge potential. This becomes clear by choosing $U'(t)=U(t)\Lambda^\dagger(t)$ with a unitary $\Lambda(t)$, we find
\begin{equation}
  -\hat{H}_1' =i \Lambda U^\dagger\partial_t (U\Lambda^\dagger)= \Lambda(-\hat{H}_1)\Lambda^\dagger+i\Lambda\partial_t\Lambda^\dagger\,,
\end{equation}
which is precisely the gauge transformation of a non-abelian gauge potential. 
Furthermore, if we project $-\hat{H}_1$ onto the subspace of  the ground states 
of $\hat{H}_2$ (or the independent sets of the corresponding graph), 
we will obtain  the non-abelian gauge matrix $A$ of the PK algorithm, namely
\begin{equation}
  -A(t)=P\hat{H}_1(t)P\,,
  \label{A}
\end{equation}
where $P$ is the projection onto the ground states of $\hat{H}_2$. If $V_{ij}\gg \| \hat{H}_1(t) \| $, 
Eq. (\ref{A}) gives an effective Hamiltonian on the subspace of $\hat{H}_2$ up to first-order. This shows that 
the Hamiltonian system governed by the gauge matrix $A$ is essentially the PXP model~\cite{Turner}, 
which is known for its quantum many-body scarring phenomenon 
and has been experimentally realized with Rydberg atom arrays \cite{Lukin2017}.

The specific correspondence between them is rather straightforward. For the original PXP model in which $\Omega(t)=\Omega$, $\Delta(t)=0$, and $\varphi(t)=0$, the PK algorithm simply sets $\theta(t)=\pi/2$ and $\phi(t)=\Omega t$. 
Quenching the PXP model to find MIS was proposed recently \cite{schiffer2024circumventing} amid with exponentially long runtime due to many-body scars. 

The above relation also offers new perspectives on the PXP model, a many-body system originally defined on a one-dimensional chain. We can now view the one-dimensional chain as the underlying graph of the PXP model, with the N\'{e}el-type state corresponding to the MIS of this simple graph, and making it possible 
to extend the PXP model general graphs. For any graph there exists a dual graph, in which each vertex represents an independent set, and each edge connects a pair of independent sets whose Hamming distance is one (see the Appendix for details). This dual graph is a median graph~\cite{Schaefer,Chepoi,wiki2sat}. The many-body PXP model, through its relation to $A$, thus can be reinterpreted 
 as a single-particle quantum diffusion over the dual graph. 
This extension has the potential to enrich the study of quantum many-body scarring. (The oscillating behavior for $\theta=\pi/2$ in Fig. 8 of Ref. \cite{wupra} can be explained in these terms.  We will discuss these ideas in more detail elsewhere.)

The relation between $A$ and $H_1$ in Eq. (\ref{A}) shows that, when the graph is fixed, the energy gap of $A$ is proportional to $-\hat{H}_1$, which suggests that to have a successful adiabatic path, it is better to let $\hat{H}_1$ have an as large energy gap as possible, in particular, during the late stage of the evolution. Fortunately, the adiabatic path in the PK algorithm indeed possesses this desirable feature. With $\hat{H}_1=(\vec{\mu}\times\vec{\mu}')\cdot\sum_j\vec{\sigma}_j$, the energy spectrum is simple and the energy gap is $\Delta E = (4\omega_{\phi}^2\sin^2(\theta/2)+\omega_\theta^2)^{1/2}$. It is easy to find
that as $\theta$ varies from $0$ to $\pi$, $\Delta E$ is steadily increasing and reaches its maximum at the end of the evolution in the PK algorithm, which are confirmed by the numerical results below.

\begin{figure}[h]
  \centering
  \includegraphics[width=0.5\textwidth]{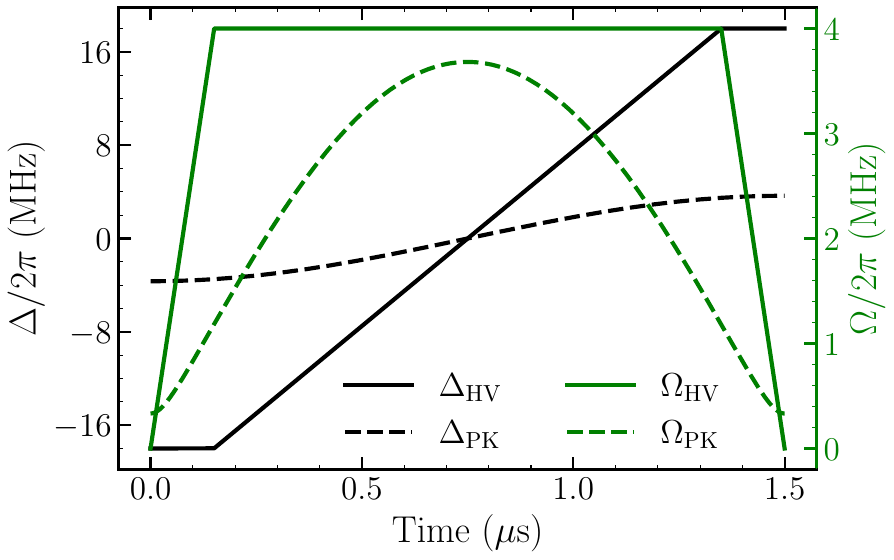}
  \caption{The unoptimized path of variational quantum adiabatic algorithm is represented by the solid lines, the same as the path given in FIG. S8. of \cite{Lukin2022}. The adiabatic path of the PK algorithm, given by Eqs. (\ref{path0}) and (\ref{path2}), is drawn as the dashed lines with
 $\omega_{\theta}=\pi/T$ and $\omega_{\phi}/\omega_{\theta}=-11$.}
  \label{fig:2}
\end{figure}

{\it Numerical results.}---We next compare numerical studies carried out with the two algorithms. They are mainly addressing two important aspects. One is the comparison of the performance of adiabatic paths, and the other is the analysis of potential accelerations in experiments adopting the PK algorithm vs the HV algorithm. The Hamiltonian $\hat{H}_2$ has many degenerate ground states, which are associated with the ISs of a graph. The minimum gap between the ISs and the excited states is $V_0$, the interaction strength between Rydberg atoms. When the changing rates $\omega_{\phi}$ and $\omega_{\theta}$ are much smaller than $V_0/\hbar$ in the PK algorithm, the system stays and evolves in the sub-Hilbert space of the degenerate ground states of $\hat{H}_2$. Its evolution in the sub-Hilbert space is governed by a non-abelian gauge matrix $A$, which has a minimum energy gap $\delta$. When  $\omega_{\theta}/\omega_{\phi}\ll \delta$, at the end of evolution $T=\pi/\omega_{\theta}$, the system displays significant amplitudes to be in states which are either MIS or its good approximations \cite{yucpl}.  For the PK algorithm to be effective, adiabatic evolution requires
\be
\label{cond}
V_0/\hbar\gg\omega_{\phi}\gg \omega_{\theta}/\delta\,. 
\ee
The explanation and physical origin of the dimensionless  energy gap  $\delta$ were given in \cite{wupra,yucpl}.  The two changing rates $\omega_{\phi}$ and 
$\omega_{\theta}$ in the PK algorithm \cite{wupra,yucpl} are related to the parameters $\Delta$, $\Omega$, and $\varphi$ in the experimental protocol \cite{Lukin2022}

\begin{subequations}
  \begin{align}
  &\Delta(t)=-2\omega_{\phi}\sin^2{\frac{\omega_{\theta}t}{2}}\,,\\
  &\varphi(t)=\arctan{\frac{\omega_{\phi}\sin\omega_{\theta}t\sin\omega_{\phi}t-\omega_{\theta}\cos\omega_{\phi}t}{\omega_{\phi}\sin\omega_{\theta}t\cos\omega_{\phi}t+\omega_{\theta}\sin\omega_{\phi}t}}\,,\\
  &\Omega(t)=\sqrt{\omega_{\phi}^2\sin^2(\omega_{\theta}t)+\omega_{\theta}^2}\,.
  \end{align}
\end{subequations}
The corresponding adiabatic path simplifies further with a single bit unitary matrix 
$V_s(t)=V(t) U_s(t)=(\vec{\mu}(t)\cdot \vec{\sigma})U_s(t)$, where $U_s=\text{diag}\{1,e^{i\phi(t)}\}$, and the simplified path is given by
\begin{subequations}
\begin{align}
\label{path0}
&\Delta(t)=\omega_{\phi} \cos(\omega_{\theta}t)\,,\\  
\label{path1}
&\varphi(t)=-\arctan\frac{\omega_{\theta}}{\omega_{\phi}\sin(\omega_{\theta}t)}+\pi\,,\\
\label{path2}
&\Omega(t)=\sqrt{\omega_{\phi}^2\sin^2(\omega_{\theta}t)+\omega_{\theta}^2}\,,
\end{align}
\end{subequations}
which is shown in Fig. \ref{fig:2} as dashed lines. Our analysis below finds that this path exhibit several advantages
over the ones in the HV algorithm discussed in Ref. \cite{Lukin2022}.

\begin{figure}[h]
  \centering
  \includegraphics[width=0.5\textwidth]{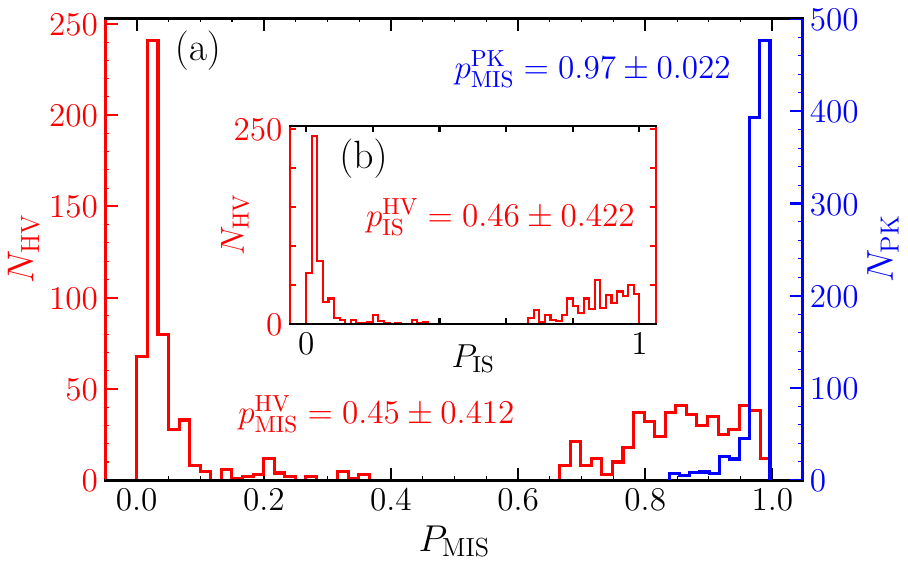}
  \caption{$P_\text{MIS}$ and $P_\text{IS}$ denote respectively the probabilities for finding MIS and IS. $N_\text{HV}$ and $N_\text{PK}$ denote the numbers of graphs employed for running HV algorithm and PK algorithms respectively. The graphs are 1000 unit disk graphs with 7 vertices. The parameters adopted are from the experiment \cite{Lukin2022}, i.e. $V_{\text{NN}}/h=107$ MHz, $V_{\text{NNN}}/h=13$ MHz, and $T=1.5$ $\mu$s. {(a)} The average success rate using the unoptimized path of HV algorithm is 45\% with standard deviation of 41.2\%. Using the adiabatic path of the PK algorithm increases the success rate to 97\% with standard deviation of 2.2\%. {(b)} The average rate of finding independent sets by HV algorithm is 46\% with standard deviation of 42.2\%, which means in most unsuccessful cases, the HV algorithm finds non-independent sets.}
  \label{fig:3}
\end{figure}

Two different variational methods are used in Ref. \cite{Lukin2022} to find an optimized evolution path using brute-force classical mathematical search over a restricted set of trial paths.  One is called the quantum approximate optimization algorithm (QAOA) \cite{QAOA} and the other is called variational quantum adiabatic algorithm \cite{VQAA}. For the QAOA, it appears that the experimental protocol (as described above) does not quite implement the optimized evolution path faithfully, which would require intermittent absence of interaction terms. For the variational quantum adiabatic algorithm, it starts with an unoptimized path shown as the solid lines in Fig. \ref{fig:2} and then the path is optimized with the help of a classical computer and experimental inputs.

The performances of the numerical results are compared for the two paths in Fig. \ref{fig:2} with the seven-vertex  graph. On average, the success rate of the PK algorithm is more than twice that of the HV algorithm. With 1000 graphs that are randomly generated, the PK algorithm has an average success rate of 97\% and the standard deviation of 2.2\%. The HV algorithm, on the other hand, shows an average success rate of 45\% and the standard deviation of 41.2\%. Further numerical results show that in most cases where the HV algorithm fails, it ends in states of non-independent sets, as shown in Fig. \ref{fig:3}(b).  

We note $\varphi(t)=0$ is chosen in our numerical calculation. To implement the PK algorithm faithfully, one actually needs to change $\varphi(t)$ according to Eq. (\ref{path1}). Numerical results show that this does not affect the performance of the PK algorithm. With $\varphi(t)$ following Eq. (\ref{path1}), we find the PK algorithm reaches an average success rate of 99\% and the standard deviation of 2.0\%.

\begin{figure}[h]
  \centering
  \includegraphics[width=0.5\textwidth]{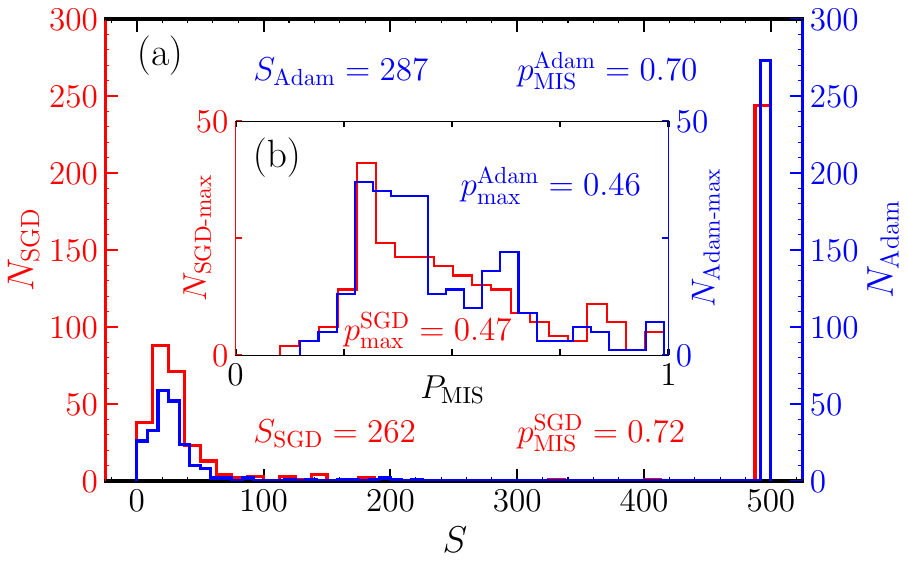}
  \caption{$S$ denotes the optimization steps required for the HV algorithm to reach the success rate $\min\ (99\% \ {\rm or}\ P_{\text{MIS}} \text{ of the PK algorithm})$ and takes a maximum value of $500$. $N_\text{SGD}$ and $N_\text{Adam}$ denote respectively the numbers of graphs using SGD or Adam optimizer. $P_\text{MIS}$ denotes the probabilities for finding MIS for the graphs. $N_\text{SGD-max}$ and $N_\text{Adam-max}$ denote respectively the numbers of graphs optimized for 500 steps using SGD or Adam optimizer. A total of 500 unit disk graphs with 7 vertices are included for each optimizer. {(a)} The average success rate and average optimization steps are 72\% and 262, 70\% and 287,  respectively from using SGD or Adam optimizer. The corresponding percentages of graphs optimized for 500 steps are 48.8\% and 54.6\% respectively.  {(b)} For the graphs that reach the maximum optimization steps, the average maximum success rates are 47\% with standard deviation of 19.1\% and 46\% with standard deviation of 17.8\%, respectively from using SGD and Adam optimizer.}
  \label{fig:4}
\end{figure}

Next, we analyze the potential acceleration of applying the PK algorithm in experiments compared with that of the HV algorithm. The HV algorithm claims that local gradient-based optimizers perform better in the experiment and that the Adam's performs the best. Thus, we employ the two widely used gradient-based optimizers, SGD and Adam. We calculate the number of optimization steps $S$ required for the HV algorithm to reach the minimum of 99\% or the PK algorithm's success rate $P_{\text{MIS}}$, with the seven-vertex graph, as shown in Fig. \ref{fig:4}. 

After 500 steps of optimization using the HV algorithm, with the use of SGD or the Adam's optimizer respectively, 48.8\% and 54.6\% of the graph still fail to reach the optimization target. The average success rate and the average optimization steps are 72\% and 262, 70\% and 287,  respectively corresponding to the use of SGD or the Adam's optimizer. The better performance of SGD is due to the fine tuning of the learning rate decay during numerical simulations. Assume that the increase in success rate is linear which represents a grossly overestimation of the optimization process, it will take at least 597 steps to reach the performance of the PK algorithm on average. Considering the maximum number of optimization steps in the experiment is about 600 and other limited optimization conditions, the optimized performance of the HV algorithm is confirmed to be much worse than that of the PK algorithm in many cases. This is explained by the fact that many classical optimization problems are difficult enough on their own as the solution space is not flat, and gradient descent can be trapped in local minima, let alone for the quantum processes applied to here.

Next we show an experimental significance of the PK algorithm that it can dramatically reduce the number of optimization steps. For a Bernoulli random variable $X$ with success probability of $p$ and sampled $m$ times, we have the Chernoff bound
\begin{eqnarray}
  \Pr(|\bar{X}-\mathrm{E}(X)|\geq \epsilon)\leq 2e^{-2m\epsilon^2}\,.
\end{eqnarray}
For Rydberg atom experiment, to measure the probability of a certain IS at the end of evolution, we denote the result differing from the IS as $X=0$ and result equal to the IS as $X=1$. The number of measurements required to acquire a $(1-\eta)$-confidence interval $[p-\epsilon,p+\epsilon]$ becomes
\begin{eqnarray}
  m\geq\frac{1}{2\epsilon^2}\log{\frac{2}{\eta}}
\end{eqnarray}
For $\epsilon=2\%$, we estimate this number is at least on the order of $10^3$. Considering that the number of variational parameters in the experiment is $\sim 10$, every step of gradient optimization requires at least $10^4$ measurements ($\sim 1.5$ hours of continuous experiment \cite{Lukin2022}). Therefore, employing the PK algorithm, at least $6\times10^6$ measurements ($\sim 900$ hours of continuous data taking) can be saved for one single graph and algorithm success rate can be improved by at least 25\% on average.

Finally we consider the effect of measurement error, assuming a measurement caused bit flip error $p_e$ for measuring an unexcited atom $\ket{0}$ to an excited state $\ket{1}$ or vice versa. The number of vertices of the graph is denoted as $n$, the number of MISs in the graph is denoted as $k$, the probability of the $i$-th MIS in the result is denoted as $p_i$, and the probability of all states with Hamming distance $j$ away from the $i$-th MIS is denoted as $p_{ij}$. Without measurement error, $p_{\text{MIS}0}=p_1+p_2+\cdots+p_k$. In the presence of the above described  measurement error, we have
\begin{eqnarray}
  \label{pmis}
  p_{\text{MIS}}=\sum_{i=1}^{k}\left[p_i(1-p_e)^n+\sum_{j=1}^{n}p_{ij}(1-p_e)^{n-j}p_e^{j}\right]\,,
\end{eqnarray}
which upon approximating to $O(p_e)$, gives
\begin{eqnarray}
  \label{error}
  p_{\text{MIS}}=(1-np_e)p_{\text{MIS}0} + p_e\sum_{i=1}^{k}\sum_{j=1}^{n}p_{ij}
\end{eqnarray}
The first term in Eq. \eqref{error} indicates that there will be a decrease in $p_{\text{MIS}}$ proportional to $n$. Assuming $p_e \sim 10^{-4}$, this result does not have a significant effect on $p_{\text{MIS}}$ when $n$ is not very large ($\sim 10^3$). But in the case of gradient optimization, the effect of the deviation of the gradient will continue to accumulate and amplify as the number of optimization steps increases. This perhaps explains why the experimental performance of the HV algorithm oscillates strongly with the number of optimization steps, while the oscillations we observe in the numerical simulations are much smaller.

We thus demonstrate that the PK algorithm, which is a fully quantum algorithm, has a more efficient and resource-saving performance than the HV algorithm which belongs to a classical-quantum hybrid algorithm. This suggests that the performance of some skillfully constructed adiabatic paths as ours is difficult to match through trivial adiabatic paths combined with brute-force searches.

\vspace{1\baselineskip}

\begin{acknowledgments}
  XJZ acknowledges discussions with Fangcheng Wang. XJZ and BW are supported by National Natural Science Foundation of China (NSFC) (Grants No.  92365202 and No.  11921005), and by Shanghai Municipal Science and Technology Major Project (Grant No.2019SHZDZX01). 
  PYG and LY are supported by NSFC (Grants
  No. 12361131576 and No. 92265205), and by the Innovation Program for Quantum Science and Technology (2021ZD0302100). 
  FW is supported by the
  U.S. Department of Energy under grant Contract 
  Number DE-SC0012567, by the European Research Council
  under grant 742104, and by the Swedish Research 
  Council under Contract No. 335-2014-7424.
\end{acknowledgments}

\def\pra{Physical Review A~}
\def\cpl{Chinese Physics Letters~}

\appendix
\section{Dual Graph}

\begin{figure}[H]
    \centering
    \begin{tikzpicture}
    \coordinate (1) at (0,0);
    \coordinate (2) at (2,0);
    \coordinate (3) at (4,0);
    \coordinate (4) at (0,-2);
    \coordinate (5) at (2,-2);
    \node [draw, fill=black, shape=circle, scale=0.3, label=above:$x_1$] at (1) {0};
    \node [draw, fill=black, shape=circle, scale=0.3, label=above:$x_2$] at (2) {0};
    \node [draw, fill=black, shape=circle, scale=0.3, label=above:$x_3$] at (3) {0};
    \node [draw, fill=black, shape=circle, scale=0.3, label=below:$x_4$] at (4) {0};
    \node [draw, fill=black, shape=circle, scale=0.3, label=below:$x_5$] at (5) {0};
    \draw (1)--(2);
    \draw (2)--(3);
    \draw (1)--(4);
    \draw (4)--(5);
    \draw (4)--(2);
    \draw (5)--(2);
    \end{tikzpicture}
    \caption{A graph with 5 vertices and 6 edges.}
    \label{fig:graph}
\end{figure}
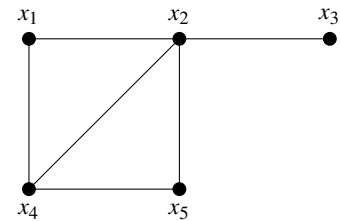

\begin{figure}[H]
  \centering

  \begin{tikzpicture}
  \coordinate (a) at (0,-0.5);
  \coordinate (b1) at (-3,-2);
  \coordinate (b2) at (-1.5,-2);
  \coordinate (b3) at (0,-2);
  \coordinate (b4) at (1.5,-2);
  \coordinate (b5) at (3,-2);
  \coordinate (c1) at (-2.25,-4);
  \coordinate (c2) at (-0.75,-4);
  \coordinate (c3) at (0.75,-4);
  \coordinate (c4) at (2.25,-4);
  \coordinate (d) at (0,-5.5);
  \draw (a)--(b1);
  \draw (a)--(b2);
  \draw (a)--(b3);
  \draw (a)--(b4);
  \draw (a)--(b5);
  \draw (b1)--(c1);
  \draw (b1)--(c2);
  \draw (b3)--(c1);
  \draw (b3)--(c3);
  \draw (b3)--(c4);
  \draw (b4)--(c3);
  \draw (b5)--(c2);
  \draw (b5)--(c4);
  \draw (d)--(c1);
  \draw (d)--(c2);
  \draw (d)--(c4);
  \node [draw, shape=rectangle,fill=white, scale=1] at (a) {$\{00000\}$};
  \node [draw, shape=rectangle,fill=white, scale=1] at (b1) {$\{10000\}$};
  \node [draw, shape=rectangle,fill=white, scale=1] at (b2) {$\{01000\}$};
  \node [draw, shape=rectangle,fill=white, scale=1] at (b3) {$\{00100\}$};
  \node [draw, shape=rectangle,fill=white, scale=1] at (b4) {$\{00010\}$};
  \node [draw, shape=rectangle,fill=white, scale=1] at (b5) {$\{00001\}$};
  \node [draw, shape=rectangle,fill=white, scale=1] at (c1) {$\{10100\}$};
  \node [draw, shape=rectangle,fill=white, scale=1] at (c2) {$\{10001\}$};
  \node [draw, shape=rectangle,fill=white, scale=1] at (c3) {$\{00110\}$};
  \node [draw, shape=rectangle,fill=white, scale=1] at (c4) {$\{00101\}$};
  \node [draw, shape=rectangle,fill=white, scale=1] at (d) {$\{10101\}$};
  \end{tikzpicture}
  \caption{The dual graph of the graph in Fig. \ref{fig:graph}. Each box (or vertex) represents an independent set. }
  \label{fig:dualgraph}
\end{figure}
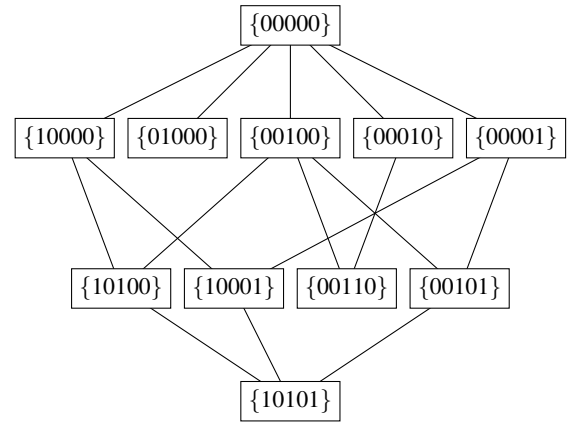

A dual graph exists for any graph, composed of all its independent sets as the vertices, and connected by edges if and only if the Hamming distance of the corresponding independent sets is one.
We use the  graph in Fig. \ref{fig:graph} as an example to illustrate dual graph. We assign each of its vertices a boolean variable. 
For this example, the five boolean variables are $x_1$, $x_2$, $x_3$, $x_4$, and $x_5$. 
Each of its independent set can then be denoted by a binary string.  For example, $\{0,0,0,0,0\}$ represents the empty set, 
$\{0,1,0,0,0\}$ represents the independent set with only one vertex $\{x_2\}$,  $\{1,0,1,0,1\}$ represents the 
maximum independent set  $\{x_1,x_3,x_5\}$, etc.  As each independent set is denoted by a binary string, 
we can define the Hamming distance between them as the Hamming distance between the corresponding binary strings. 
For example, the Hamming distance between the empty set and the set with one vertex $\{x_2\}$ is one. 
the Hamming distance between the empty set and the MIS  $\{x_1,x_3,x_5\}$ is three.

The graph of Fig. \ref{fig:graph} has 11 independent sets,  that are  shown in Fig. \ref{fig:dualgraph} as marked boxes, or vertices of the dual graph that are connected by an edge between a pair of  vertices if their Hamming distance is one. For any graph, its dual graph is  a median graph.

\end{document}